# Spontaneous emission enhancement from polymer-embedded colloidal PbS nanocrystals into Si-based photonics at telecom wavelengths

M. Humer, R. Guider, W. Jantsch, T. Fromherz

Institute of Semiconductor and Solid State Physics, Johannes Kepler Universität,

Altenbergerstrasse 69, A-4040 Linz, Austria.

Abstract:

We experimentally demonstrate the coupling of optically excited PbS nanocrystal (NC) photoluminescence (PL) into Si-based ring resonators and waveguides at 300K. The PbS NCs are dissolved into Novolak polymer at various concentrations and applied by drop-casting. The coupling mechanism and the spontaneous emission enhancement are experimentally investigated and compared to theoretical predictions. Quality (Q) factors of 2500 were obtained in emission and transmission for wavelengths centered around 1.45µm. PL intensity shows a linear dependence on the excitation power and no degradation of the Q factors. Devices with stable optical properties are obtained by this versatile technique.

Manuscript:

For the near-infrared spectral range, lead based colloidal nanocrystal quantum dots (NCs) are of considerable interest due to their unique optical and electronic properties[1,2,3]. Their absorption and emission wavelengths depend on the dot size, which can be easily controlled during chemical synthesis. Therefore, they are very attractive as building blocks for nanophotonic applications at telecommunication wavelengths. Especially Lead Sulfide (PbS) NCs have found a wide range of potential applications such as surface emitting lasers[4], polymer strip-loaded plasmonic waveguides[5] and solar cells[6].

A convenient way to manufacture stable and reproducible quantum dot based devices is to introduce the NCs into polymers like PMMA, SU-8 or PFCB[7]. In fact, by the choice of proper ligands, the quantum dots can be mixed into various solvents of organic polymers where, after solvent evaporation, the polymer acts as a matrix in which the NCs are homogeneously distributed[8,9]. The choice of a polymer structurable by photo-, electron-beam- or soft-lithography opens the possibility to fabricate photonic devices from this compound material. Furthermore, inorganic surface capping of colloidal nanocrystals was achieved recently, allowing the encapsulation of NCs into an amorphous $As_2S_3$ matrix, which resulted in an all-inorganic thin film with stable infrared luminescence in the telecom region[10].

Previous studies of PbS NCs coupled to silicon based photonic crystal microcavities have either relied on free-space or fiber coupled micro-photoluminescence methods to pump and collect the emission from high-$Q$ cavities[11,12,13,14]. Lasing from colloidal NCs has also been shown for example for CdSe NCs in the visible spectral range[15] or for PbS NCs in the infrared[16], where in both cases microcapillary tubes were used as resonators and filled with quantum dots.

In this letter, we deposit PbS NCs on top of all-integrated Si-based ring resonators coupled to bus waveguides and report on the coupling of the emission from PbS quantum dots to the resonator modes measured at the output of a Si bus waveguide at room temperature. In particular, we perform studies on the stability of the emission and investigate the coupling mechanism and the device performance as a function of the amount of quantum dots and of the excitation power.

Commercially available PbS colloidal nanocrystals (NCs) from Evident Technologies (EviDot core LNR03LPB) with a diameter of 5.1nm and a peak PL emission at 1385nm, dissolved in toluene (40mg/ml) are used as optically active material for the polymer-NC blend. The PL spectrum of a spin cast pure NC film after toluene evaporation is shown in Fig. 1a by the dotted line. As polymer host material Allresist AR-N 7700.18 (Novolak based negative electron beam resist) is used and NC-polymer blends containing different nanocrystal concentrations varying from 0.5vol% to 10vol% are prepared and analyzed.

For testing the degradation of the NC luminescence, the polymer-NC blend is drop cast onto a silicon substrate and cured on a hotplate at 75°C for 1min. The PL spectra of the

NC-polymer films are measured by exciting the blends with a frequency doubled Nd:YAG laser at 532nm wavelength. In Fig. 1a, the normalized PL spectra of the pure NCs (dotted line) and the polymer-NC blend (full lines with superimposed symbols) are shown for different aging steps of the blend (squares: 67 days after drop casting). For a quantitative monitoring of the PL intensity's temporal evolution, it is essential to keep both the number of NCs within the detection volume of the PL setup and the excitation intensity constant for measurements at subsequent observation intervals. In an integrated optical setup as sketched in Fig. 1c, in which a drop of polymer-NC solution is deposited into the opening of an SU-8 passivation layer, only a fixed number of NCs in close vicinity to the waveguide can emit their PL radiation into the waveguide. In Fig. 1b, the PL peak intensity of this fixed number of NCs measured at the end of the waveguide is plotted for several observation intervals. Figure 1b shows the excellent stability of the composite material's PL emission intensity over a period of 37 days. Because of this temporal and spectral stability shown in Figs. 1a and 1b, the Novolak-NC composite material was used as active medium in the experiments discussed in the following.

For defining ring resonators coupled to bus waveguides, the 220nm thick, semi-insulating ($<10^{15}$ cm$^{-3}$ Boron doped) device layer of an SOI wafer was processed into 600nm wide and 220nm high Si monomode waveguides by Amo Ltd. using electron beam lithography and ICP-RIE etching. The 20μm diameter ring resonators are separated from the bus waveguides by a 150 nm wide gap.

The samples are first passivated by spin casting a 500nm thick MicroChem SU-8 2000.5 resist. Next, UV lithography was used to open L=60μm≈2πr and W=30μm wide windows centered around the ring resonators, where r is the radius of the ring (see Fig. 1c for a sketch). This relation between r and L was chosen to ensure equal interaction surfaces of bus waveguide and resonator with the Novolak-NC blend. The windows were filled by drop casting the blend and subsequent solvent evaporation on a hotplate at 75°C for 1 minute.

The refractive indices of the Novolak AR-N 7700.18 (n=1.575) and the SU-8 2000.5 (n=1.572) differ only slightly[17,18] around 1.5µm wavelength. Therefore, the interface between the SU-8 passivation and the NC loaded Novolak film in the SU-8 -windows does not cause significant mode distortion in the underlying bus waveguide. Additionally, the Novolak film is easily removed by Allresist AR 300-70 remover, whereas the hardened SU-8 remains unaffected by this remover. Compared to previously reported approaches[19], this high selectivity of the remover allows us to completely replace Novolak-NC films with different NC concentration in a simple way without affecting the optical properties of ring-resonators and bus waveguides Therefore, the same resonator can be used for analyzing the spontaneous emission rate (SER) enhancement with respect to the NC concentration of the Novolak film.

Transmission measurements are performed using a fiber-coupled continuous wave InGaAsP laser source tunable from 1460nm to 1580nm with an output power adjusted to

2mW. To inject and collect light, polarization maintaining single mode tapered fibers are used. Light detection is performed by a fiber-coupled InGaAs detector.

For investigating the PL emitted into the resonator and waveguide modes, the excitation lasers (532nm or 657nm wavelength) are focused onto the devices. PL emission is collected at the end of the bus waveguide and dispersed by a fiber-coupled monochromator (resolution 0.2nm).

Figure 2a shows a TM polarized photoluminescence spectrum of the ring resonator – bus waveguide device covered with a Novolak film containing 4vol% PbS NCs .For obtaining the results shown in Fig. 2a, an excitation laser (532nm wavelength) with 80 µm spot diameter (larger than the SU-8 window dimensions) was used. Thus, NCs all over the window area are excited and can emit PL radiation into the continuous bus-waveguide or discrete ring resonator mode spectrum, provided that they are located within the respective region of the evanescent electric field surrounding the Si waveguides. A contour plot of the in-plane component of the waveguide mode's electric field obtained by a 3D Beam Propagation Method (BPM) simulation is shown in Fig. 2c. Accordingly, the PL spectrum measured at the end facet of the bus waveguide consists of a broadband PL signal superimposed by narrow peaks of intensity (see Fig. 2a). Figure 2b shows that in transmission and emission experiments the resonances are observed at identical wavelengths and with similar peak widths.

In Figure 3a, the ratio between the measured resonator peak- and bus waveguide-PL intensity (described in Fig. 2a) is labeled as intensity ratio $R = I_{res}/I_{wvg}$ and shown for NC concentrations in the range between 0.5 and 10vol%. For these measurements a constant excitation intensity of 32W/cm$^2$ was used.

For a further analysis of R and its dependence on the NC concentration of the active material we note that the peak resonance intensity $I_{res}$ is proportional to the number of participating NC emitters $N_{res}$, to the SER into the resonator modes $SE_{res}$ and the collection efficiency between the ring and the waveguide η ($I_{res} = N_{res}*SE_{res}*\eta$)[20]. The PL intensity measured for the bus waveguide can be written as $I_{wvg}= N_{wvg}*SE_{wvg}$. Due to our choice of L~ 2πr (see Fig. 1c for a definition of L and r), the number of emitters coupled to the bus waveguide $N_{wvg}$ approximately equals $N_{res}$. Therefore, $R = (SE_{res} / SE_{wvg}) * \eta$ depends on the NC concentration only via the collection efficiency η.

According to the definition of the Purcell factor[21,22], $SE_{res}/SE_{wvg}$ equals $F_{res}/F_{wvg}$ where $F_{res}$ and $F_{wvg}$ denote the Purcell factors of ring resonator and bus waveguide, respectively, which can be calculated following Ref.[23]. Thus, knowledge of η allows a comparison between our experimental results and theoretical predictions of the SER enhancement in ring resonators. According to Ref[24], $\eta=Q_t/Q_c$, where $Q_t$ denotes the total quality (Q) factor of a resonator coupled to a read out waveguide and $1/Q_c=1/Q_t - 1/Q_i$. Here, $Q_c$ describes the coupling between the bus waveguide and the ring and $Q_i$ is related to the intrinsic and parasitic losses of the resonator[25,26].

From waveguide transmission experiments as shown in Fig. 2b, $Q_c$ can be determined using $Q_c=2Q_t/(1\pm\sqrt{T_{min}})$ [25,26], where $T_{min}$ is the minimum in the normalized transmission spectrum at resonance and the ± corresponds to the over- and under- coupled loading condition[27]. From the same transmission experiments $Q_t$ can be evaluated as ratio of the resonance wavelength and the full -3dB bandwidth of the resonance. Thus a decomposition of $Q_t$ into $Q_c$ and $Q_t$ as shown in Fig. 3c is obtained.

For NC concentrations below 4vol%, $Q_i$ is approximately constant, but $Q_c$ decays with increasing NC concentration, indicating an approach towards the critical coupling as a consequence of an increase of the refractive index of the composite material with the NC concentration. This increased coupling is also the reason for the increase of R shown in Fig. 3a. For NC concentrations >4vol%, $Q_i$ decays with increasing NC concentration, indicating increased scattering losses as the NC act as scattering centers. These scattering losses are the main reason for the saturation and decay of R shown for NC concentrations >4vol% in Fig. 3a.

As shown in Fig. 3c, similar total Q factors are observed in transmission and emission experiments for all NC concentrations investigated in this work. Assuming therefore that also $Q_c$ is equal for transmission and emission, η can be determined from the Q values obtained from the transmission spectra and used for experimentally evaluating the Purcell factor ratio of ring resonator and bus waveguide R/η as discussed above. This enhancement is plotted in Fig. 3b vs the NC concentration (full line). For NC concentration <6vol%, a 13 times larger spontaneous emission rate into the ring resonator modes as compared to the bus waveguide modes is obtained.

Also shown in Fig. 3b by the broken line is the ratio between the Purcell enhancement in a ring resonator $F_{res}$ and of a straight waveguide with equal cross sections. Following Ref.[23], $F_{res}/F_{wvg} = SE_{res}/SE_{wvg} = (1/\pi)(\lambda/n_{eff})(Q_i/L)$, where $\lambda=1.45\mu m$, $n_{eff}=2.14$ is obtained from 3D BPM simulations, L is the length of our cavity and $Q_i$ is the intrinsic quality factor taken from our experimental analysis.

The calculated enhancement is in very good agreement with the theoretical analysis. The deviation at higher concentrations (in the region of high coupling) is explained by the fact that coupling of NC emission from the bus waveguide into the resonator modes results in dips in the bus waveguide PL spectrum, which are not taken into account in our evaluation of $I_{res}$.

Furthermore, the PL intensity measured at the output of the bus waveguide was studied as a function of the continuous wave pump laser intensity for the devices containing NC concentrations of 4vol% and 6vol%. For these experiments an AlGaInP laser operating at 657nm was focused to a spot size slightly larger than the ring resonator diameter (~20µm), resulting in a negligible fraction of the bus waveguide being illuminated. Therefore, Fig. 4b contains just the ring resonator peaks. For both concentrations, an approximately linear increase of the PL output power with the pumping intensity was observed for the total range of excitation intensities accessible in this work [see Fig. 4a for devices covered by a Novolak film with the optimal NC concentration of 4vol%]. Neither a degradation of the Q factors nor a saturation of the output power with increasing intensity is observed within our current experimental capabilities.

Finally, we want to note that the NC integration method developed in this work is directly applicable to resonator structures that provide larger mode overlap with the active material, like for example vertically slotted ring resonators[26].

In conclusion, a convenient and reliable way of coupling the emission of PbS NCs into the resonating modes of SOI ring resonators on an all-integrated optoelectronic chip at telecommunication wavelengths is demonstrated. The influence of the nanocrystal concentration in the Novolak polymer host material on the performance of the device is studied and a 13-fold enhancement due to the cavity was demonstrated experimentally, in good agreement with theoretically predicted values. An optimal compromise for maximal out coupled intensity is found for a nanocrystal concentration of around 4vol%. Performance is also studied with respect to excitation power and a linear behavior is observed. Novolak-NC blends show unaltered performance weeks after preparation, indicating the high stability of this composite material. As the emission wavelength of chemically synthesized nanocrystals can be precisely controlled, the working bandwidth of nanocrystal based optoelectronic chips can be easily tuned to other wavelength regions within the transparency region of the Si waveguides.

Acknowledgements:

This work was supported by the Austrian Nanoinitiative within the PLATON project cluster under project numbers 819654, 819655, 819656, 834913, and 834924. We also acknowledge W. Heiss and G. Bauer for fruitful discussions.

Figures:

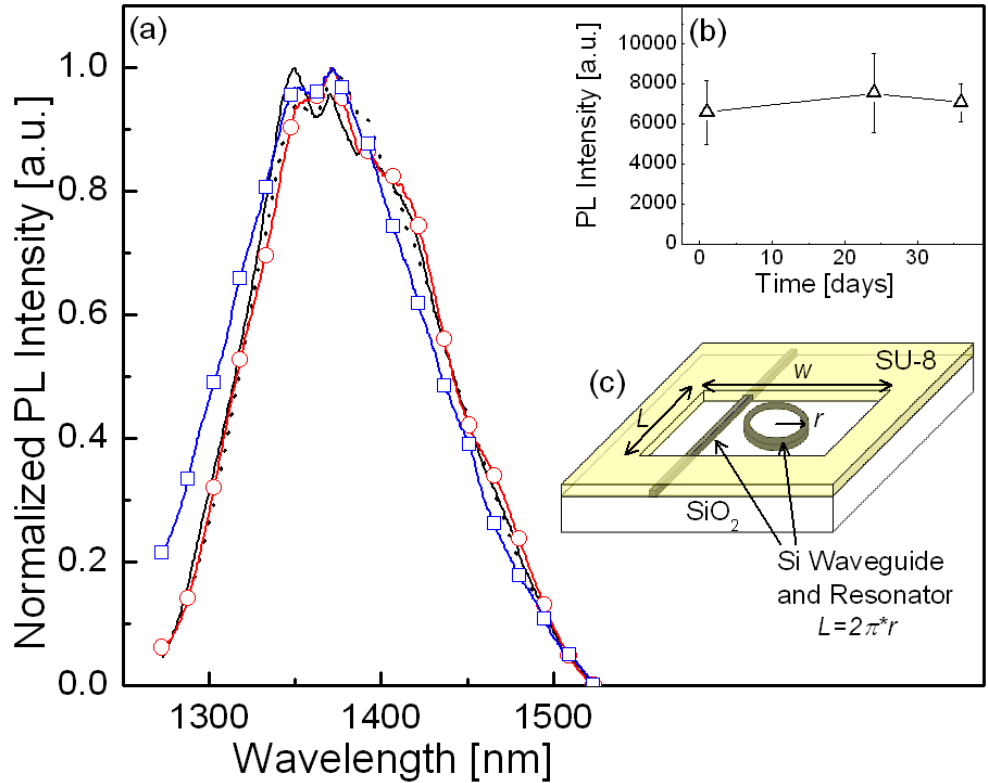

Figure 1 (a): Normalized room temperature PL spectra of pure PbS NCs drop cast onto a Silicon substrate (dotted line). The solid lines correspond to the spectra of the NC blend after drop casting (black) and 7 (red, circles) and 63 (blue, squares) days after drop casting.

Figure 1 (b): Intensity of room temperature PL of a defined amount of the NC blend taken under similar excitation in function of the days after drop casting.

Figure 1 (c): Sketch of the investigated ring resonator coupled to the silicon waveguide.

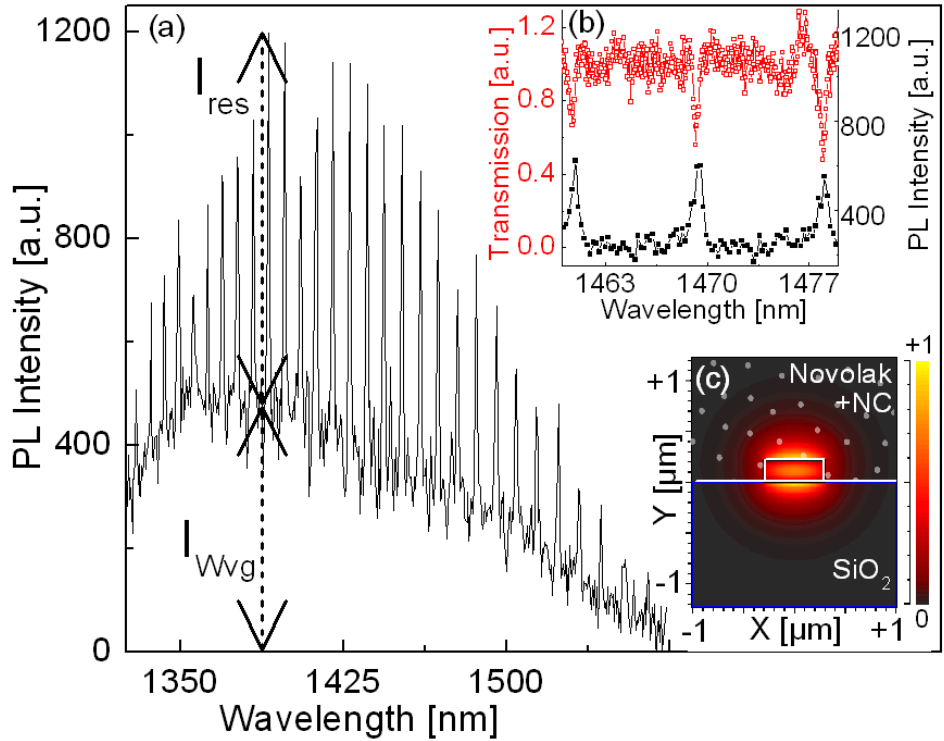

Figure 2 (a): Room temperature PL spectrum of Novolak containing 4 vol% of PbS drop cast onto the ring resonator collected at one output facet of the waveguide. Intensities of the ring resonance PL and the bus waveguide PL are indicated as $I_{res}$ and $I_{wvg}$.

Figure 2 (b): Confrontation between the transmission measurements and the PL emission measurements.

Figure 2 (c): 3D Beam Propagation Method (BPM) simulation of the silicon waveguide mode (TM polarization, $\lambda_{freespace}=1.45\mu m$) indicating the spatial overlap with the active material on top of the waveguide (indicated by the dotted area).

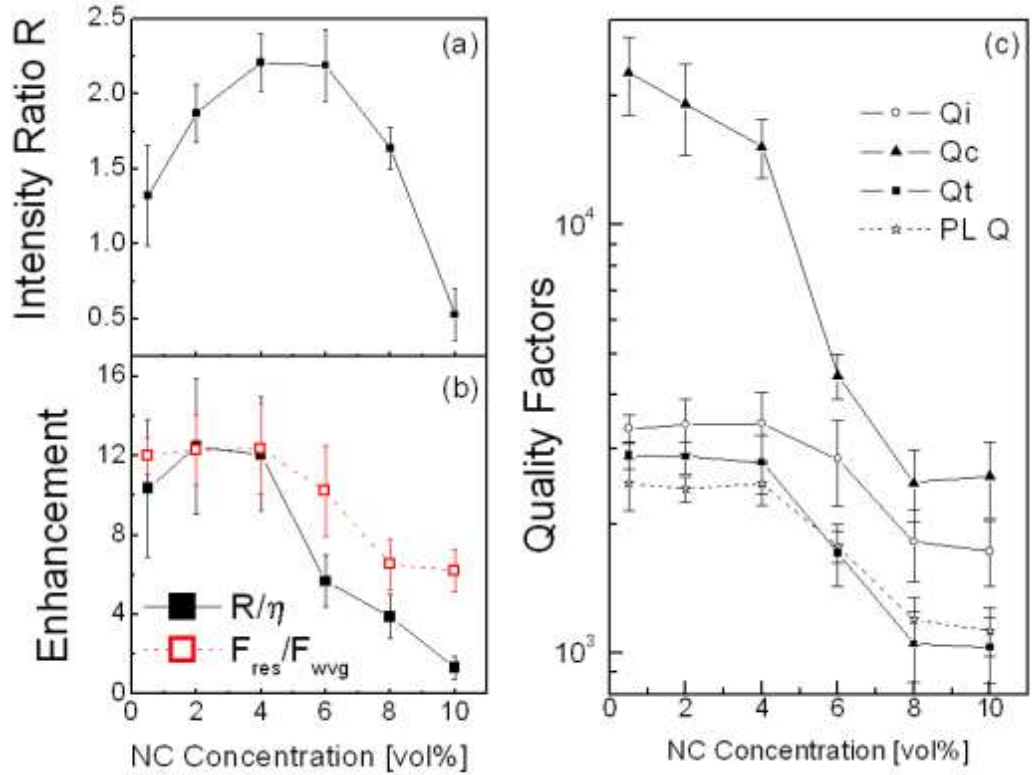

Figure 3 (a): Average Intensity ratio R as a function of the NC concentration, defined as the ratio between the PL of the resonating modes and the PL background of the bus waveguide.

Figure 3 (b): Measured enhancement factors, defined as R/η (full line, squares), and theoretically estimated enhancement $F_{res}/F_{wvg}$ (dashed red line, open squares) as a function of the nanocrystal concentration.

Figure 3 (c): Istrinsic, Coupling and Total Q factors ($Q_i$, $Q_r$ and $Q_t$ respectively) calculated from the measured transmission spectra and Q factors calculated from the obtained PL spectra (PL Q).

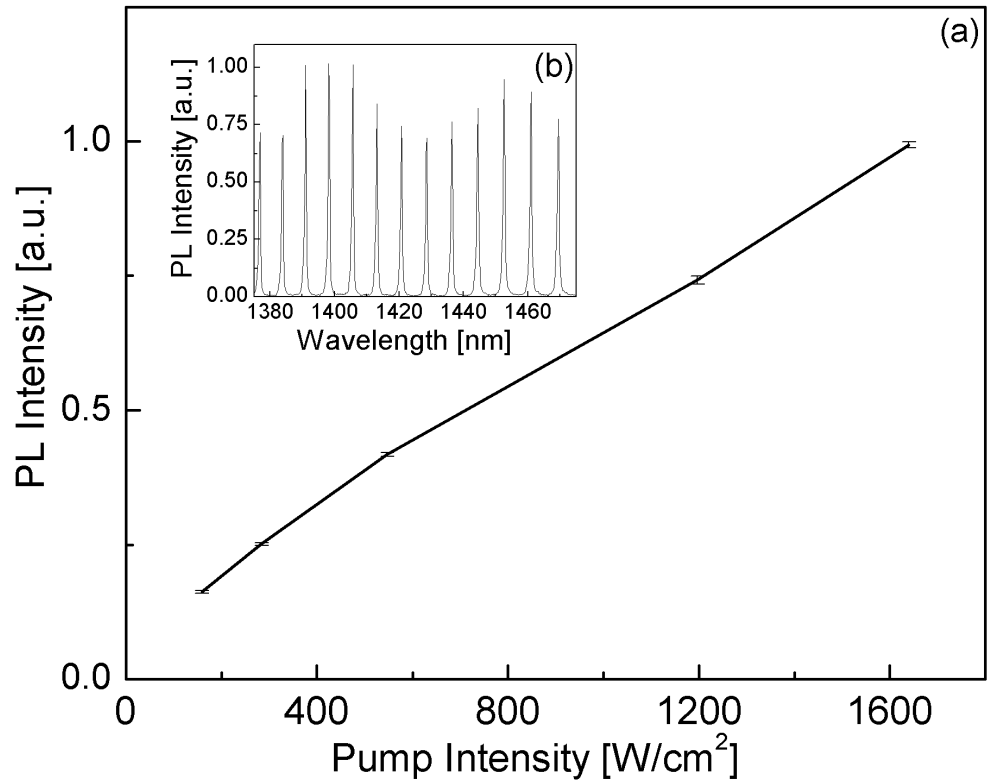

Figure 4 (a): Excitation power dependent PL intensities detected at the end facet of the bus waveguide of Novolak containing 4vol% of PbS drop cast onto the ring resonator.

Figure 4 (b): Room temperature PL spectrum at a pump intensity of 1.64kW/cm$^2$.